\documentclass[11pt,singlecolumn,showpacs,preprintnumbers,amsmath,amssymb]{revtex4}
\usepackage{graphicx}% Include figure files
\usepackage{dcolumn}% Align table columns on decimal point
\usepackage{bm}% bold math

\def\be{\begin{equation}}
\def\ee{\end{equation}}
\def\ba{\begin{array}}
\def\ea{\end{array}}
\def\bea{\begin{eqnarray}}
\def\eea{\end{eqnarray}}

\begin{document}
\preprint{}
%%%%%%%%%%%%%%%%%%%%%%%%%%%%%%%%%%%%
\title{\large\bf On the balance energy and nuclear dynamics\\
in peripheral heavy-ion collisions}

\author{Rajiv Chugh}

\author{Rajeev K. Puri}
\email{rkpuri@pu.ac.in}
\affiliation{\it Department of Physics, Panjab University, \\
Chandigarh -160 014, India.\\ }

\begin{abstract}
We present here the system size dependence of balance energy for
semi-central and peripheral collisions using quantum molecular
dynamics model. For this study, the reactions of
$Ne^{20}$+$Ne^{20}$, $Ca^{40}$+$Ca^{40}$, $Ni^{58}$+$Ni^{58}$,
$Nb^{93}$+$Nb^{93}$, $Xe^{131}$+$Xe^{131}$ and
$Au^{197}$+$Au^{197}$ are simulated at different incident energies
and impact parameters. A hard equation of state along with
nucleon-nucleon cross-sections between 40 - 55 mb explains the
data nicely. Interestingly, balance energy follows a power law
$\propto{A^{\tau}}$ for the mass dependence at all colliding
geometries. The power factor $\tau$ is close to $-\frac{1}{3}$ in
central collisions whereas it is $-\frac{2}{3}$ for peripheral
collisions suggesting stronger system size dependence at
peripheral geometries. This also suggests that in the absence of
momentum dependent interactions, Coulomb's interaction plays an
exceedingly significant role. These results are further analyzed
for nuclear dynamics at the balance point.
\end{abstract}

\pacs{25.70.Pq}
 \maketitle

\section{Introduction}
Heavy-ion collision at intermediate energies is a complex
phenomenon that depends crucially on the interplay of the mean
field and nucleon-nucleon binary collisions. The dominance of
either of these ingredients can shed light  on the properties of
nuclear matter at the extreme of temperature and density. Among
various observables and non-observables in heavy-ion collisions,
directed collective flow is considered to be a very important
observable due to its extreme sensitivity towards the model
ingredients and physics involved \cite{herr,gust,ren}. The very
same observable has also been robust for the understanding of
nuclear equation of state - a question that is still open and
extensively debated in the literature
\cite{hart,and,grein,risc,pan,zhou1,sull}. The directed collective
flow has been reported by many authors to be positive at higher
incident energies whereas same is deeply negative at low incident
energies. While going from the low to high incident energies,
there comes a typical point where no preference is observed for
the flow, therefore, it disappears. This particular incident
energy where attractive and repulsive interactions balance each
other (and flow disappears) is termed as balance energy or
$E_{bal}$ \cite{zhou1,west1,krof1,sunil,ogli,krof,he,mag,sood2}.
At the balance point, scatterings due to the attractive mean field
occurring at negative angles are counter balanced by the repulsive
action of the binary nucleon-nucleon collisions occur at positive
angles \cite{krof1,krof,sood,soff}.
\par
The composite dependence of the $E_{bal}$ on the mean field
equation of state (eos) and nucleon-nucleon cross-section
($\sigma_{NN}$) can be sorted out by noticing the sensitivity of
$E_{bal}$ on the system size as well as on the impact parameter of
the reaction. In this connection, very recently, Puri and Sood
\cite{sood}, conducted a very detailed and exhaustive study on the
energy of vanishing flow over entire periodic table with masses
between 20 and 394. Their study had shed light on various aspects
of nuclear dynamics. Unfortunately, this study along with all
other studies reported in the literature are limited to
central/semi-central collisions only \cite{sood}. A very little
attention has seen paid in the literature for semi-central and/or
peripheral collisions, where physics is derived by the low density
dynamics \cite{sull,ogli,sood,soff,buta,moli,pak}. Among various
attempts involving non-central impact parameters, Magestro et. al.
\cite{mag} reported the results of disappearance of flow over
large impact parameters. Their study, however, was limited to
heavier systems only. One should keep in the mind that the
dynamics in lighter nuclei is entirely different than that of
heavier colliding nuclei. It remains, therefore, a challenging
question to see how directed transverse flow behaves at
perpheral/non-central collisions. We plan to address this question
in this study.
\par
In an another study, Puri et.al. \cite{sunil} extracted the
strength of nucleon-nucleon cross-section for the reaction of
$Zn^{64}+Al^{27}$ at different collision geometries. They firmly
indicated that for central collisions, a cross-section of 40 mb
was good enough, whereas an enhanced value of cross-section was
needed for peripheral collisions. It still remains to be seen
whether this observation holds good over entire periodic table or
not.
\par
Here we plan to understand the mass dependence of disappearance of
flow in these low excited geometries. We shall also attempt to
parameterize the balance energy for the system size effects. The
present study is conducted within the framework of quantum
molecular dynamics (QMD) model, which is discussed in section II.
In section III, we discuss the results and summary is presented in
section IV.
\par

\section{The Model}

In QMD model \cite{aich,qmd1,qmd2}, each nucleon, represented by a
Gaussian distribution, propagates with classical equations of
motion as:
\begin{equation}
\frac{d{\bf r}_i}{dt} = \frac{d H}{d{\bf p}_i},
\end{equation}
\begin{equation}
\frac{d{\bf p}_i}{dt} = - \frac{d H}{d{\bf r}_i},
\end{equation}
where Hamiltonian $H$ is given by :
\begin{equation}
H = \sum_i \frac{{\bf p}_i^2}{2m_i} + V^{tot},
\end{equation}
with
\begin{equation}
V^{tot} = V^{loc} + V^{Yuk} + V^{Coul}.
\end{equation}
Here  $V^{loc}, V^{Yuk}$ and $V^{Coul}$  represent, respectively,
the local two- and three-body Skyrme, Yukawa and Coulomb
interactions. Under the local density approximation, Skyrme part
of the nucleon interaction can be written as:
\begin{equation}
V^{loc}=\alpha/2\sum_i\rho_i+\frac{\beta}{\gamma+1}\sum_i\rho_i^{\gamma}.
\end{equation}
Here $\rho_i$ is the nucleon density. The coefficients $\alpha$
and $\beta$ give proper ground state properties and $\gamma$ is
used to generate different compressibilities. The different values
of compressibility ($\gamma$) give possibility to look for the
role of different equations of state termed as soft ($\gamma$ =
{200} MeV) and stiff ($\gamma$ = {380} MeV) equations of state.
The values of $\alpha$ and $\beta$ for soft and hard equation of
state are -0.356 GeV, 0.303 GeV and -0.124 GeV, 0.0705 GeV
\cite{aich}. Following many studies \cite{krof,he,sood2,sood}
listed in the literature, we also use a hard equation of state.
\par
When propagating nucleons come too close to each other, they can
collide elastically or inelastically depending upon the available
center of mass energy. The collision probability depends on the
cross section $\sigma_{NN}$. Several different forms of
$\sigma_{NN}$ are available in the literature. In many studies
\cite{sood2,sood}, one has taken constant value of $\sigma_{NN}$.
Generally it is assumed to be between 20 mb and 55 mb. The
strength of nucleon-nucleon cross-section will be extracted by
simulating various reactions with $\sigma_{NN}$ between 40 and 55
mb. The above QMD model has been shown to be very useful in many
observables/non-observables \cite{qmd1,qmd2}.

\section{Results and Discussion}
\par

\begin{figure}
\centering
\includegraphics[width=12cm]{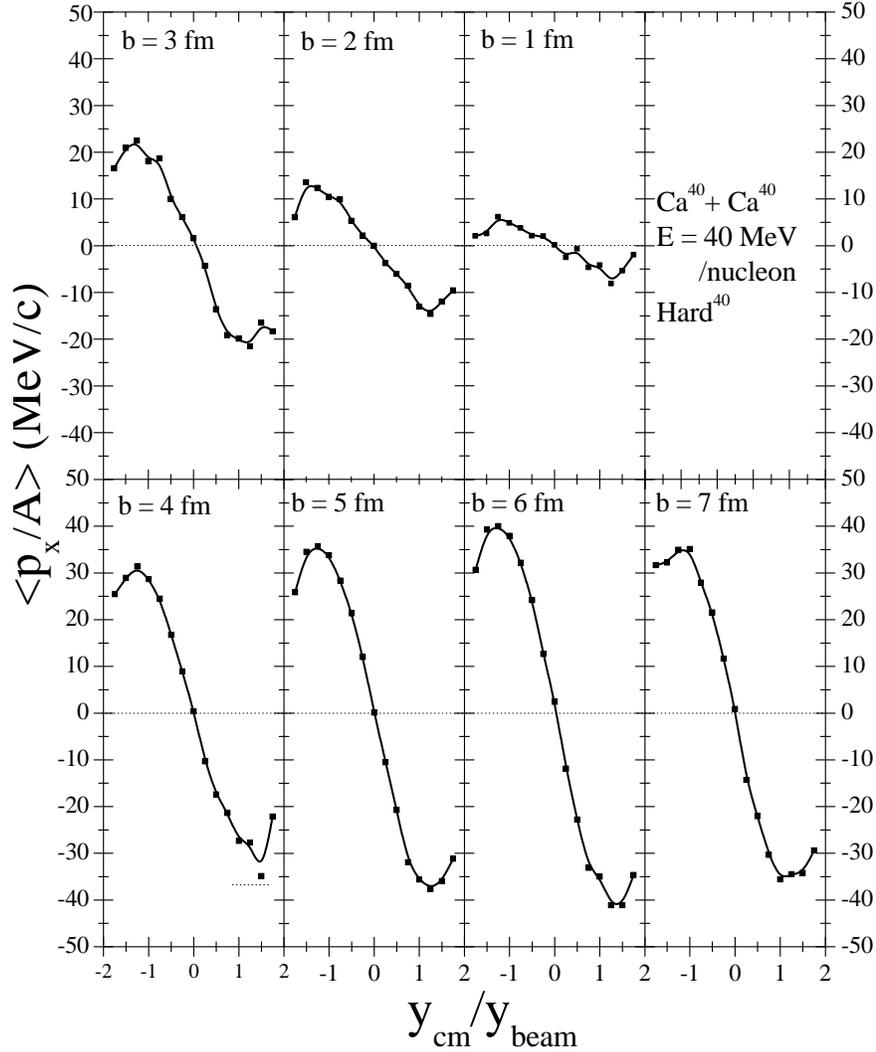}
%\vskip - 1.0cm
\caption{Average $<p_{x}/A>$ (MeV/c) as a function of scaled
rapidity $y_{cm}/y_{beam}$. We display the results (anticlockwise)
at impact parameters b = 1, 2, 3, 4, 5, 6 and 7 fm.}
\end{figure}
\par
\begin{figure}
\centering
\includegraphics[width=12cm]{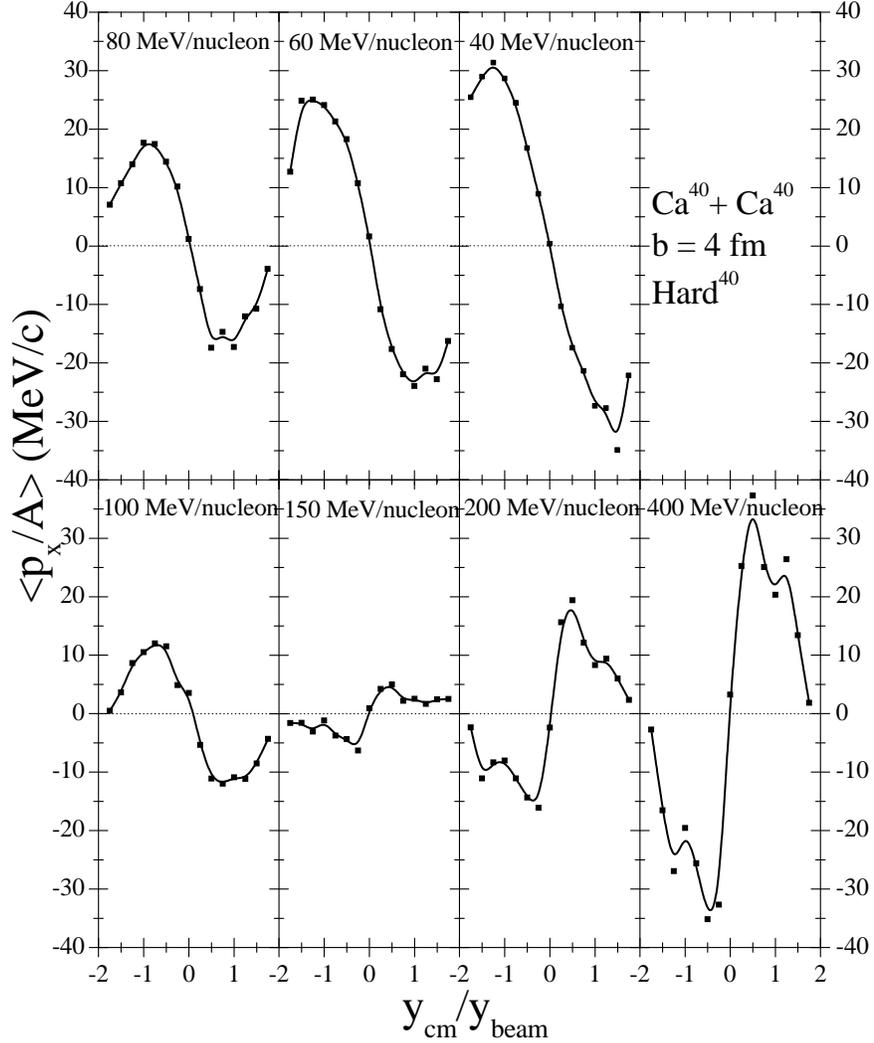}
%\vskip - 1.0cm
\caption{Same as in fig. 1 but here we display the results
(anticlockwise) at energies 40, 60, 80, 100, 150, 200 and 400
MeV/nucleon.}
\end{figure}
\par

\begin{figure}
\centering
\includegraphics[width=8cm]{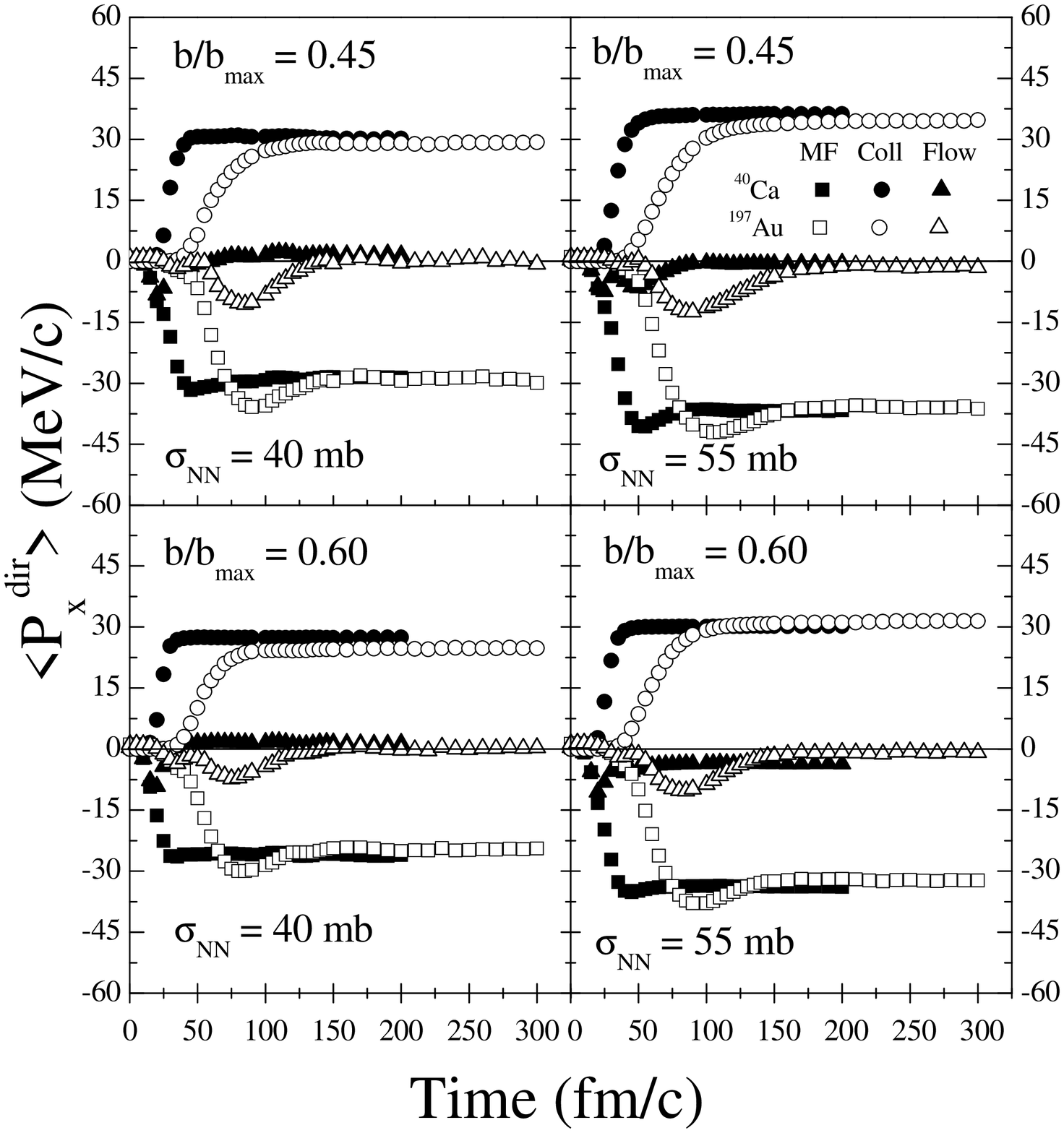}
\vskip - 1.0cm \caption{The time evolution of total directed
transverse momentum $<P_{x}^{dir}>$ along with mean field and
collision parts at the energy of vanishing flow. Solid symbols
represent $Ca^{40}$ whereas open symbols are for $Au^{197}$.}
\end{figure}

\par
For the present study, symmetric reactions of $Ne^{20}+Ne^{20},
Ca^{40}+Ca^{40}, Ni^{58}+Ni^{58}, Nb^{93}+Nb^{93},
Xe^{131}+Xe^{131}$ and $Au^{197}+Au^{197}$ are taken. The entire
colliding geometry is considered with main emphasis on the
semi-central as well as on the peripheral collisions. Throughout
the present analysis, a hard equation of state along with
$\sigma_{NN}$ = 40 - 55 mb has been used. In fig.1, we display the
changes in the transverse momentum $<p_{x}/A>$ as a function of
the rapidity distribution $y_{cm}/y_{beam}$. Here we vary impact
parameter from b = 1 to b = 7 fm at a step of unity. We see that
for b = 1 fm, flow is nearly isotropic and slope is zero. The
negative slope increases as we move from central to peripheral
collisions. This demonstrates that for larger impact parameters,
matter is attractive in the presence of reduced number of
collisions. It is worth that the slope of the $<p_{x}/A>$ depends
crucially on the incident energy. In fig. 2, we display the
$<p_{x}/A>$ for the same reaction as fig. 1 but at b = 4 fm and
for different incident energies. We see that slope is negative at
small incident energies whereas it turns less negative with the
increase in the incident energies. At particular energy (between
100 and 150 MeV/nucleon), we see a change in the slope from
negative to positive values. This slope is steeper at higher
incident energies.
\par
In Fig.3, the transverse directed momentum, $<P_{x}^{dir}>$ is
plotted as a function of reaction time for the collisions of
$Ca^{40}$+$Ca^{40}$ and $Au^{197}$+$Au^{197}$ using $\sigma_{NN}$
= 40 and 55 mb, respectively. The displayed results are at reduced
impact parameters $b/b_{max}$ ($\hat{b}$) = 0.45 and 0.60. The
plots are displayed at the energy of vanishing flow i.e. balance
energy. For a better understanding, we divide the total transverse
momentum into contribution resulting from the mean field and from
the collision parts. From the figure, one observes that the
transverse momentum due to mean field (representing squares) is
being equated by that due to binary nucleon-nucleon collisions
(circles), resulting in net zero flow (triangles) at the end of
the reaction. The mean field contribution dominates the reaction
during early evolution resulting in negative (attractive)
transverse flow. It is worth mentioning that if mean field
dominates, the transverse flow will be negative whereas the
dominance of nucleon-nucleon collisions will result in to a
positive flow. At ultra low incident energies, the phenomena
emerging due to mean field alone are dominant \cite{malik}. The
reaction persists longer for heavier systems (up to 100 fm/c or so
for $Au^{197})$ compared to lighter colliding nuclei (which is
about 50 fm/c). This happens due to lower energy of vanishing of
flow in heavier systems compared to light systems. The different
values of the flow due to mean field using different $\sigma_{NN}$
happens due to different energies of vanishing flow.
\par
\begin{figure}
\centering
\includegraphics[width=8cm]{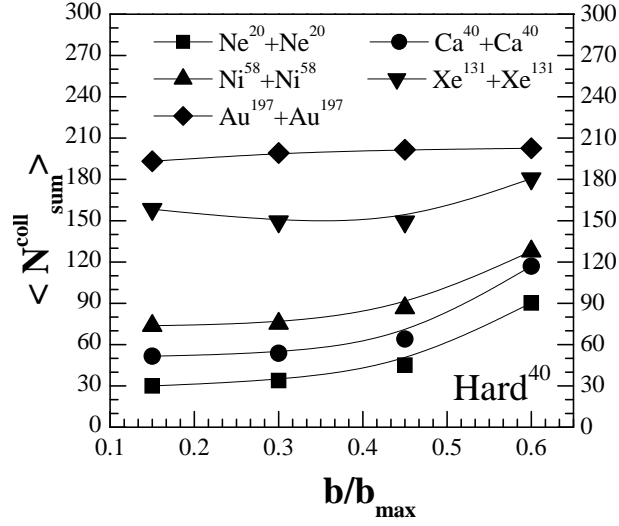}
\vskip - 2.0cm \caption{The total binary collisions
($<N^{coll}_{sum}>$) at 200 fm/c as a function of reduced impact
parameter ($\hat{b}$). Lines are drawn to guide the eye. Here the
reactions of $Ne^{20}$+$Ne^{20}$, $Ca^{40}$+$Ca^{40}$,
$Ni^{58}$+$Ni^{58}$, $Xe^{131}$+$Xe^{131}$ and
$Au^{197}$+$Au^{197}$ are taken.}
\end{figure}

\par

\begin{figure}
\centering
\includegraphics[width=8cm]{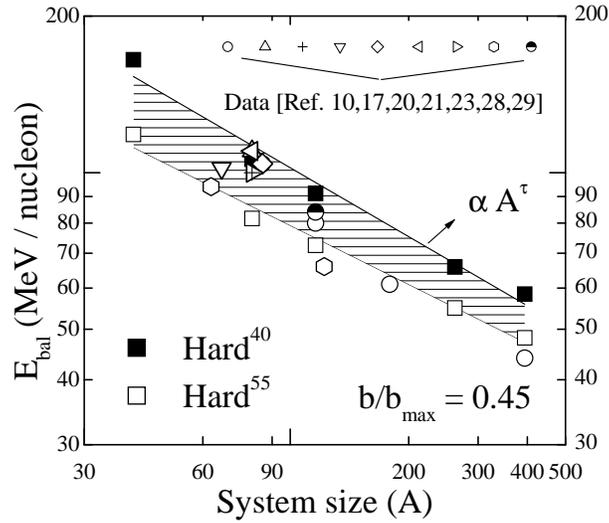}
\vskip - 2.0cm \caption{The balance energy as a function of system
mass at $\hat{b}$ = 0.45. The data points are taken from
references mentioned in text. Shaded region represents the energy
of vanishing flow obtained with $\sigma_{NN}$ = 40 and 55 mb}
\end{figure}
\par

\begin{figure}
\centering
\includegraphics[width=8cm]{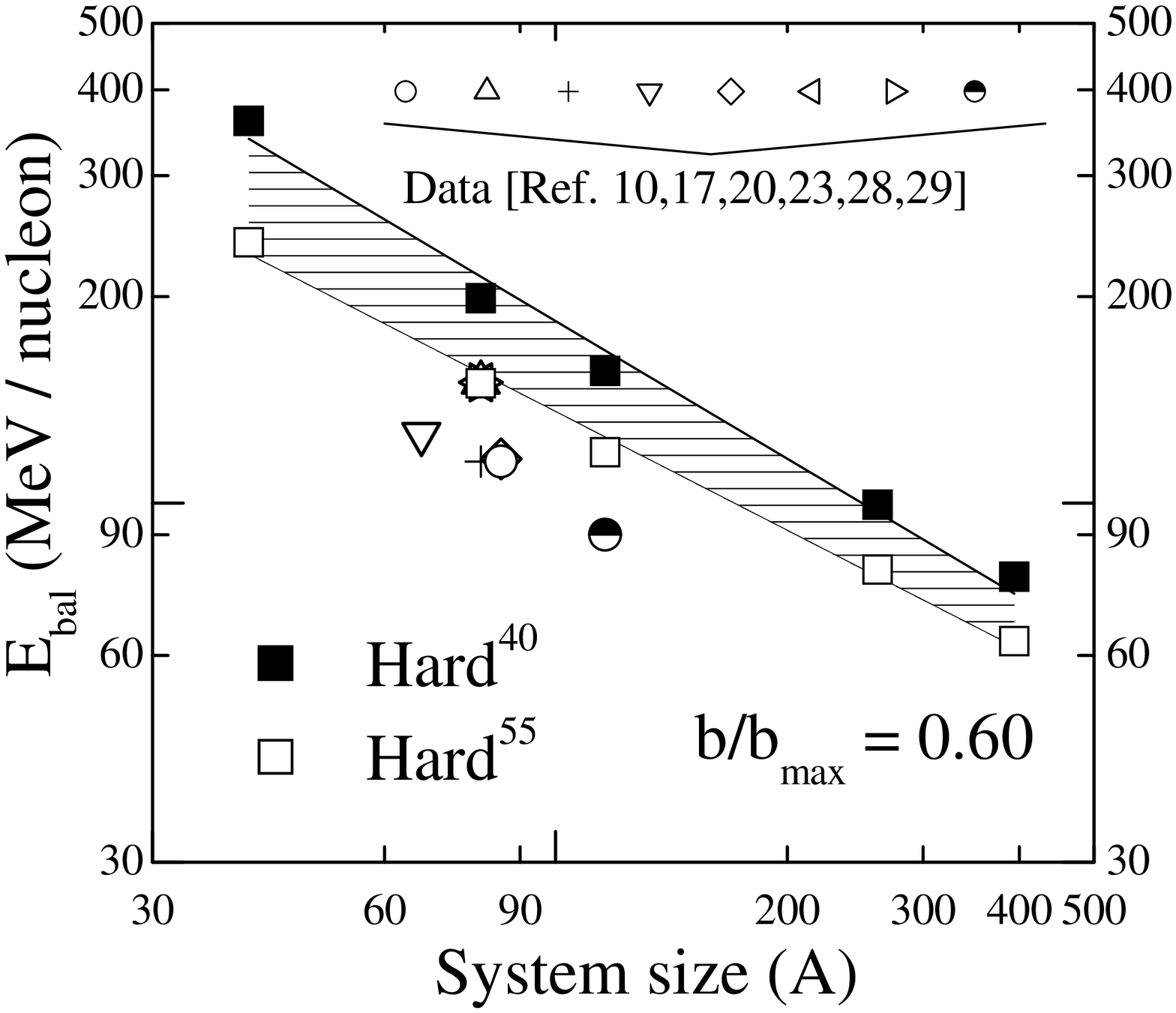}
\vskip - 2.0cm \caption{Same as fig.5, but at $\hat{b}$ = 0.60.}
\end{figure}

\par
We plot in Fig.4, the number of allowed collisions
$<N^{coll}_{sum}>$, for various systems as a function of reduced
impact parameter $\hat{b}$. This dependence of binary collisions
on impact parameter yields several interesting aspects: For the
lighter masses, there is a marked enhancement in the binary
collisions at peripheral collisions. This happens due to
exceedingly high balance energy at these impact parameters in
lighter colliding nuclei. For example, the $E_{bal}$ for
$Ne^{20}$+$Ne^{20}$ at $\hat{b}$ = 0.15 is 76 MeV/nucleon, which
goes up to 360 MeV/nucleon for $\hat{b}$ = 0.60. On the other
hand, binary collisions for heavier colliding nuclei (like
$Au^{197}$ + $Au^{197}$) are almost independent of the impact
parameter. This is due to the fact that the energy of vanishing
flow is very small in $Au^{197}$+$Au^{197}$ reaction (between 40
and 78 MeV/nucleon) and at these  low incident energies, most of
the binary collisions are blocked, and as a result no significant
impact parameter dependence is obtained for binary collisions in
heavy-ion systems.
\par
\begin{figure}
\centering
\includegraphics[width=8cm]{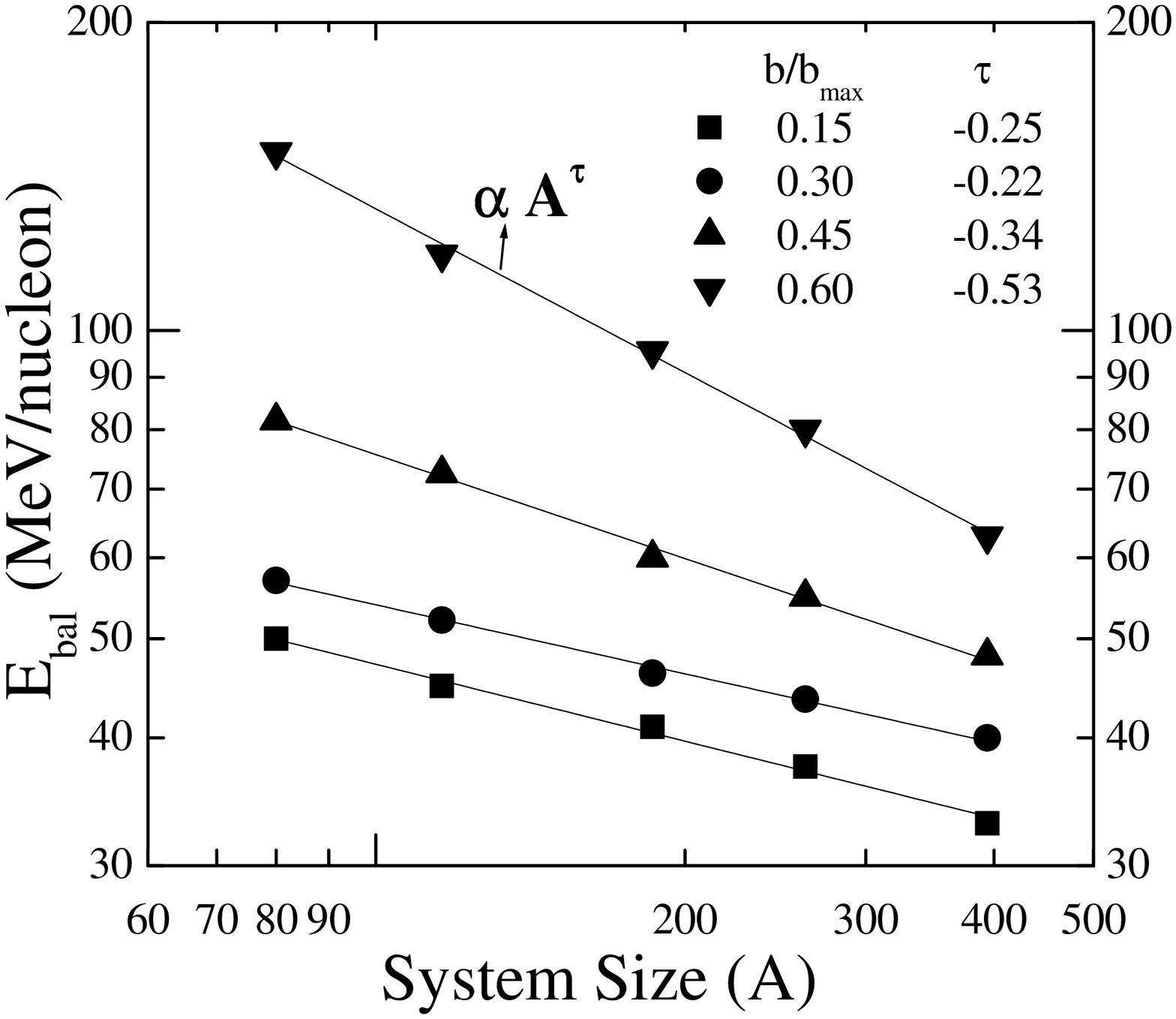}
\vskip -2.0cm \caption{The balance energy $E_{bal}$ against system
mass at different reduced impact parameters. The displayed results
are using QMD model with hard EOS at 55 mb cross-section.}
\end{figure}
\par
\begin{figure}
\centering
\includegraphics[width=12cm]{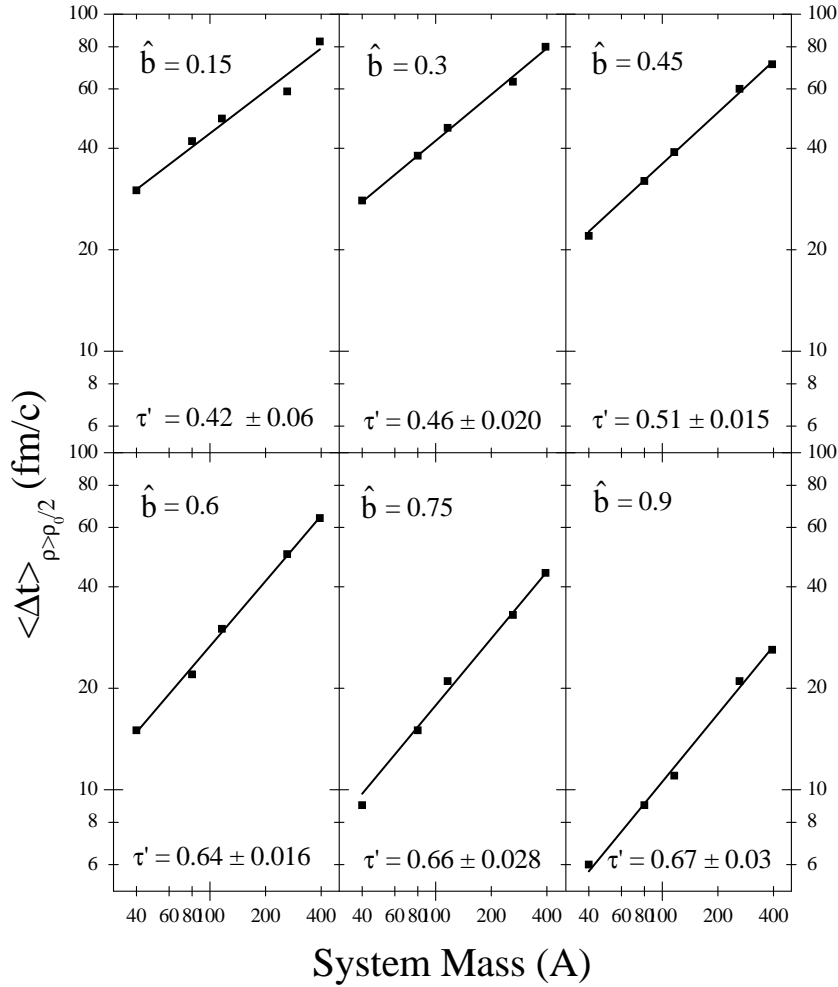}
\vskip - 0.0cm \caption{The time zone for which density remains
higher than half of the normal matter versus system size for
different impact parameters.}
\end{figure}
\par
\begin{figure}
\centering
\includegraphics[width=12cm]{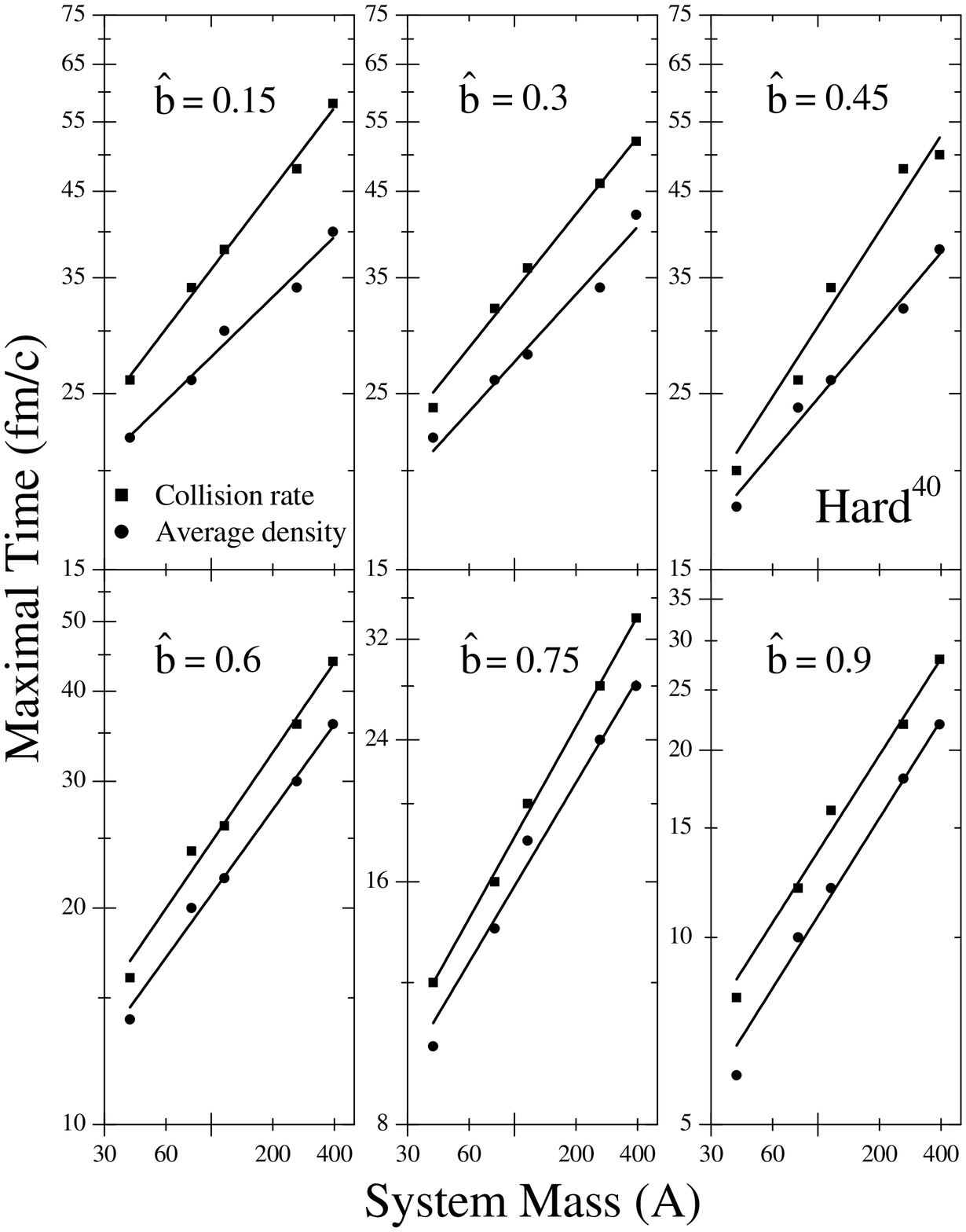}
\vskip - 0.0cm \caption{Maximal time for collision rate and
average density as function of system size.}
\end{figure}
\par
In figs.5 and 6, we plot the energy of vanishing flow as a
function of total system mass $A$ (mass of projectile + mass of
target) for entire mass range between 40 and 394 units.
Remarkably, our present calculations with $\sigma_{NN}$ between 40
and 55 mb explain experimental data for $\hat{b}$ = 0.45 nicely,
whereas some deviation appears in the case of $\hat{b}$ = 0.60.
The data shown in the figure are for the reactions of
$Ar^{40}+Sc^{45}, Ca^{40}+Ca^{40}, Ni^{58}+Ni^{58},
Fe^{58}+Fe^{58}, Kr^{86}+Nb^{93}, Ar^{36}+Al^{27},
Zn^{64}+Ni^{58}, Zn^{64}+Ti^{48}$ and $Au^{197}+Au^{197}$
\cite{sull,mag,soff,buta,pak,pak1,west}. The scarcity of data at
larger impact parameters restricts the similar detailed analysis.
We notice that different cross-sections have larger influence for
light colliding nuclei that reduces significantly for heavier
systems. The cause suggests the fact that the balance energy for
heavier system is much smaller compared to lighter nuclei. As a
result, cross-section plays an insignificant role in heavier
systems. This mass dependence can be parameterized by a power law
of the form $\propto{A^{\tau}}$. As we see, the $\tau$ for
$\hat{b}$ = 0.45 is -0.44 and -0.37 for our calculations with
$\sigma_{NN}$ = 40 and 55 mb respectively, whereas it is -0.66 and
-0.57 for $\sigma_{NN}$ = 40 and 55 mb at $\hat{b}$ = 0.60. The
deviation of our calculated energy of vanishing flow from
experimental data suggests that the dynamics of low energy domain
needs further investigation. We also notice that around A = 100
units, no clear mass dependence appears. This could also be due to
the fact that impact parameter in these measurements is not fixed.
It varies with reaction. The impact parameter variation can also
have effect on the energy of vanishing flow.
\par
Fig.7 shows our calculated balance energy versus system mass at
reduced impact parameters of 0.15, 0.30, 0.45 and 0.60. All
calculated points in the energy scale are fitted with a power law
$\propto {A^{\tau}}$. One notice that the slope of plots increases
with impact parameter. The values of power factor $\tau$ are
-0.25, -0.22, -0.34 and -0.53 respectively, for $\hat{b}$ = 0.15,
0.30, 0.45 and 0.60 (Note that in this figure and fitting,
$Ne^{20}$+$Ne^{20}$ reaction has been excluded, therefore slope
deviates slightly compared to fig.6).We see that the value of
$\tau$ depends strongly on the colliding geometry. For instance,
for central collisions, $E_{bal}$ is 50-57 MeV/nucleon for
$Ca^{40}+Ca^{40}$ reaction, whereas it increases to 149.5
MeV/nucleon for $\hat{b}$ = 0.60.
\par
For central collisions, we observe that the power factor is close
to (-1/3). This is a well known trend reported by number of
authors earlier \cite{west1,sood2,sood} and has been very well
justified in the literature in terms of interplay between the mean
field and binary collisions. The increase in the slope of
$E_{bal}$ with impact parameter can be due to failure of static
equation of state to reproduce transverse motion of particles. The
effect is more severe for lighter nuclei. This is partially due to
the fact that with weaker binary nucleon-nucleon collisions,
Coulomb's repulsion plays a decisive role. This role is very
strong in heavier nuclei and at peripheral collisions, therefore
making slope steeper. It still remains to be seen how momentum
dependent interactions can alter these findings.
\par
To further understand this similarity in the balance energy
obtained at different impact parameters, we display in fig.8, the
evolution time for which density remains higher than half of the
matter density i.e. the time for which $\rho > 0.5\rho_{0}$. This
gives an important information about the interaction time at
different impact parameters. We display the results at $\hat{b}$ =
0.15, $\hat{b}$ = 0.3, $\hat{b}$ = 0.45, $\hat{b}$ = 0.6,
$\hat{b}$ = 0.75 and $\hat{b}$ = 0.9. Interestingly, we see that
in all cases, time zone for different masses can be explained with
the help of a power law $\propto{c'A^{\tau'}}$. The factor $\tau'$
increases systematically from central to peripheral reactions. It
is 0.42 for central whereas it increases to 0.67 for $\hat{b}$ =
0.9. A careful look reveals that the effect is less for heavier
nuclei (for Au+Au it is 80 fm/c for $\hat{b}$ = 0.15 whereas, it
is 20 at $\hat{b}$ = 0.9). On the other hand, it has drastic
effect for lighter one where it slashes to $10\%$ of central
values. This is the cause why balance energy changes drastically
for lighter nuclei compared to heavier ones and we have different
slopes compared to peripheral ones.
\par
In fig.9, we display the maximal time of average density and
collision rate for the same systems as shown in fig.8. Here we
also see again a similar trend as seen in fig.8. There is a
monotonous increase in the slope from central to peripheral
collisions.
%%%%%%%%%%%%%%%%%%%%%%%%%%%%%%%%%%
\section {Summary}
\par
In the present work, we analyzed the disappearance of flow (or
alternately the balance energy) for semi-central and peripheral
collisions using quantum molecular dynamics model. The present
study was undertaken by using a hard equation of state along with
constant cross-sections between 40 and 55 mb. We see a mass
dependence $\propto A^{-1/3}$ for central collisions, whereas this
dependence becomes steeper with impact parameter. This deviation
from -1/3 trend suggests exceeding role of Coulomb's repulsion in
low density and excited situations. Nuclear dynamics at the
balance point indicates drastic changes in lighter system compared
to heavy one in agreement with balance energy.
\par
This work is supported by Department of Science and Technology,
Government of India.
%%%%%%%%%%%%%%%%%%%%%%%%%%%%%%%%%%%%%%%%%%%%%%%%%%%%%%%%%%%%%%%%%%%%%%%

\end{document}